\documentclass[a4paper,11pt]{article}
\usepackage{jcappub} 
\usepackage{aas_macros}
\title{\boldmath New axion bounds derived from the 100-parsec {\em Gaia} DR3 white dwarf luminosity function}  
\author[a,b]{M. L. Alberino,}
\author[a,b]{M. M. Miller Bertolami,}
\author[c]{M. E. Camisassa,}
\author[d, e, f]{A. Caputo}
\author[c]{and S. Torres}
\affiliation[a]{Instituto de Astrofísica de La Plata, Consejo Nacional de Investigaciones Cientíﬁcas y Técnicas Avenida Centenario (Paseo del Bosque) S/N, B1900FWA La Plata, Argentina}
\affiliation[b]{Facultad de Ciencias Astronómicas y Geofísicas, Universidad Nacional de La Plata Avenida Centenario (Paseo del Bosque) S/N,
B1900FWA La Plata, Argentina}
\affiliation[c]{Departament de Física, Universitat Politècnica de Catalunya, \\ c/Esteve Terrades 5, 08860 Castelldefels, Spain}
\affiliation[d]{Department of Theoretical Physics, CERN, Esplanade des Particules 1, P.O. Box 1211, Geneva 23, Switzerland}
\affiliation[e]{Dipartimento di Fisica, ``Sapienza'' Universit\`a di Roma \& Sezione INFN Roma1, Piazzale Aldo Moro
5, 00185, Roma, Italy}
\affiliation[f]{Department of Particle Physics and Astrophysics, Weizmann Institute of Science, Rehovot 7610001, Israel}
\emailAdd{mlalberino@fcaglp.unlp.edu.ar}
\abstract{
The axion, a well-motivated hypothetical particle arising in extensions of the Standard Model,  can be produced copiously within the hot, compact cores of white dwarf  stars.
The shape of the white dwarf luminosity function (WDLF) is a powerful tool for constraining theoretical particles that would imply an additional cooling channel in white dwarfs. In this work, and for the first time, we use the 100-parsec {\em Gaia} DR3 white dwarf sample and compare it with theoretical predictions. 
We have simulated synthetic populations of white dwarfs using a population synthesis code based on Monte Carlo techniques, incorporating realistic observational errors, and based on state-of-the-art white dwarf models that incorporate the anomalous cooling caused by the presence of  axions.  Axion bremsstrahlung emission rates were implemented using the latest theoretical calculations.
We find that, for the brightest white dwarfs in the sample ($M_{\mathrm{Bol}}<10$), the $\chi^2$ statistic is largely insensitive to the assumed stellar formation rate (SFR), which is typically the dominant uncertainty in modeling the Galactic-disk WDLF.  The resulting $\chi^2$ analysis disfavors a sizable additional cooling contribution. This conclusion contrasts with earlier studies in which axion–electron couplings in the range $0.7\times10^{-13}<g_{ae}<2.1\times10^{-13}$ provided mildly improved fits to the Galactic-disk WDLF. We attribute the discrepancy to simplifying assumptions in previous modeling and to the substantially improved observational quality of the 100-pc {\em Gaia} DR3 sample. We obtain the upper limit $g_{ae} < 1.68\times10^{-13}$ ($95\%$ C.L.), which is among the strongest available.
}
\begin{document}
\maketitle
\flushbottom
\section{Introduction}
\label{sec:intro}
The axion was originally proposed as a compelling dynamical solution to the strong-CP problem in QCD
\cite{PecceiQuinn, Weinberg,Wilczek}, emerging as a pseudo-Nambu–Goldstone boson of a spontaneously 
broken global symmetry. Beyond this minimal motivation, a broad class of axion-like particles (ALPs) arises 
naturally in well-motivated extensions of the Standard Model, including many string-inspired constructions
\cite{Svrcek:2006yi, Arvanitaki:2009fg}, making light pseudoscalars a generic target for both particle-physics and astrophysical probes. Soon after its proposal it was realized that stellar astrophysics was an excellent tool to constrain axion properties \cite{1978JETPL..27..502V}. Over roughly the past two decades, the axion field has experienced a major 
acceleration in both theoretical development and experimental activity. On the theory side, this includes 
increasingly systematic treatments of axion/ALP effective theories, production mechanisms, and their 
phenomenology across laboratory, astrophysical, and cosmological settings. On the experimental side, a 
diverse global program has formed, spanning haloscope searches for Galactic dark-matter axions, 
helioscopes targeting solar axions, purely laboratory “light-shining-through-walls” techniques, and a growing 
portfolio of novel detection concepts exploiting electromagnetic, nuclear, and condensed-matter technologies. 
Comprehensive overviews of this rapidly evolving landscape can be found in recent reviews~\cite{Irastorza:2018dyq, DiLuzio:2020wdo,Chadhaday,Semertzidis, OHare:2024nmr,Caputo2024}. A powerful and largely complementary 
avenue comes from stellar evolution. If axions couple to Standard Model particles, they can be produced 
efficiently in stellar interiors and escape, providing an additional energy-loss channel that alters cooling 
times and, consequently, various stellar observables~ \cite{Caputo2024,Carenza2025,Raffelt_book}. White dwarfs  are particularly attractive 
laboratories in this context: their late-time evolution is essentially a cooling process, 
and the white dwarf luminosity function (WDLF) is especially sensitive to the cooling rate.
This connection has motivated a long line of studies using white dwarfs—via both WDLF measurements \cite{isern2008,miller2014, isern2018} and white dwarf 
pulsation properties \cite{1992ApJ...392L..23I,2012corsico,2012corsico_b, 2016corsico, 2016battich}—to test axion-induced cooling. 
In particular, the WDLF provides a powerful probe of light pseudoscalars with an electron coupling. In white dwarfs, these particles would be emitted predominantly through a Bremsstrahlung process, and would act as a non-negligible extra-cooling for the brightest and youngest white dwarfs \cite{isern2008,miller2014,isern2018}.

In the specific case of the QCD axion, the axion--electron coupling arises in concrete ultraviolet completions such as the DFSZ model \cite{Zhitnitsky:1980tq,Dine:1981rt}. In DFSZ realizations one may express the coupling as
\begin{equation}
g_{ae} = 2.8 \times 10^{-14}\, \left(\frac{m_a}{\mathrm{meV}}\right)\cos^2\beta \, ,
\label{eq:dfsz}
\end{equation}
where $m_a$ is the axion mass and $\cos^2\beta$ is a model-dependent parameter. For generic ALPs, instead, $g_{ae}$ and the ALP mass are independent parameters, and for a given axion mass the WDLF constrains $g_{ae}$ directly.

Using WDLF determinations based on white-dwarf samples from the Sloan Digital Sky Survey (SDSS) and the SuperCOSMOS Sky Survey (SSS),  \cite{miller2014} reported an upper bound $g_{ae} < 2.3 \times 10^{-13}$ at 95\% confidence level. Moreover, analyses of these earlier WDLFs \cite{miller2014,isern2018} found a modest improvement in the fit for axion masses in the range $2.5 < m_a \cos^2\beta < 7.5~\mathrm{meV}$.
It is also worth noting that the WDLF employed in these analyses are based on magnitude-limited samples, using the $1/V_{max}$ method \citep{1968ApJ...151..393S} to account for the incompleteness of the sample at low luminosities \citep{2006AJ....131..571H,2008AJ....135....1D}. In this method, each star is counted weighted by $1/V_{max}$, where $V_{max}$ is the maximum volume where the star would still be visible considering the magnitude cut of the sample. This way, fainter stars contribute more to the WDLF than brighter stars. Although this method has been widely used in past decades, recent advances in the observed population of white dwarfs have allowed the construction of volume-limited samples. Indeed, the {\em Gaia} space mission has provided a 100 pc volume-limited sample that reaches $\sim 94\%$ for a 
parallax relative error lower than  $\sim 10 \%$ \citep{Jimenez2018} . 
Once a proper classification of the distinct Galactic populations \cite{2019MNRAS.485.5573T} and spectral classifications are performed \cite{JE2023,2023Garcia-Zamora,2024Vincent} this sample allows the construction of a WDLF exempt from typical observational biases present in magnitude-limited samples \citep[see e.g.][]{JE2023,2024NewAR..9901705T}. Furthermore,
these previous analyses of axion properties using WDLFs adopted proper-motion selection methods, while {\em Gaia} parallax measurements have enabled the determination of distances and absolute magnitudes for these stars. Also, new classification techniques based on Random Forest algorithms that use the full parameter space of the observations are now available to separate those white dwarfs that belong to the Galactic thin disk \citep{2019MNRAS.485.5573T}. 

Owing to unaccounted observational biases and modeling assumptions in the construction of old theoretical WDLFs, previous constraints on axion properties are likely overestimated. Moreover,
recent revisions of axion emissivities \cite{bottaro} indicate that the emission rates previously reported in the literature \cite{nakagawa1,nakagawa2} were also slightly overestimated. The high quality of the 100 pc volume limited white dwarf sample of the {\em Gaia} DR3 sample \cite{JE2023} opens the possibility of improving the quality and robustness of previous constraints. Achieving this goal, however, requires a faithful propagation of observational uncertainties in the comparison between data and theory.
The best way to achieve this is by means of  Monte Carlo population-synthesis simulations, in which one generates synthetic white dwarf populations and constructs the corresponding WDLF while folding in the {\em Gaia} 
performance characteristics (astrometric and photometric uncertainties, completeness, and selection function) in a forward-modeling framework.
Moreover a proper set of cooling white dwarf models in which axion emission is computed self-consistently with the thermal structure of the stars, following the approach advocated by \cite{miller2014}, is desirable. In the present work we implement this strategy and apply it to the {\em Gaia}~DR3 100~pc sample. 

The paper is organized as follows: In Section \ref{sec:tools}, we discuss the axion emission rates and the computation of stellar models. In Section \ref{sec:wdlf}
the generation of synthetic populations. In Section \ref{sec:impact},
we quantitatively assess the impact of axions on the theoretical WDLF and provide statistical tests to constrain $g_{ae}$ at the 95\% confidence level.
Finally, in Section \ref{sec:results}, we summarize our results and conclusions and outline possible directions for future work.
\section{White dwarf models and input physics}\label{sec:tools}
\subsection{Standard white dwarf physics}
The stellar evolution computations presented on this work were carried out with the ${\tt LPCODE}$ stellar evolution code, widely used in the study of different problems related to the evolution of white dwarfs \cite{2005A&A...435..631A,2010ApJ...717..183R, 2013ApJ...775L..22M}. A detailed description of the code is available at \cite{2010ApJ...717..183R} and references therein. 
The equation of state for the high- and low-density regimes are taken from \cite{1994ApJ...434..641S}  and \cite{1979A&A....72..134M}, respectively.
Conductive opacities are those of \cite{2007ApJ...661.1094C} while radiative
opacities are those of the OPAL project \citep{1996ApJ...464..943I}
complemented at low temperatures by the molecular opacities produced by
\cite{2005ApJ...623..585F}. 
White dwarf models computed with ${\tt LPCODE}$ also include detailed non-gray model atmospheres to provide accurate boundary conditions for our models, which include non ideal effects in the gas equation of state, see \cite{2012A&A...546A.119R} for details. Neutrino cooling by
Bremsstrahlung, photo and pair production are included following the recipes
of \cite{1996ApJS..102..411I}, while plasma processes are included according
to \cite{1994ApJ...425..222H}. All relevant energy sources are taken into
account in the simulations, including residual nuclear burning, the release of
latent heat and the gravitational energy associated with the phase separation
in the carbon-oxygen profile induced by crystallization. The inclusion of all
these energy sources is done self-consistently and locally coupled to the full
set of equations of stellar evolution.
In addition, time-dependent element diffusion during white dwarf evolution is taken into account following the treatment of multicomponent gases of \cite{1969fecg.book.....B} and the fully implicit numerical approach presented in \cite{2020A&A...633A..20A}, including thermal, chemical, and gravitational diffusion; and taking into account the effects of Coulomb interactions on the gravitational settling \citep{2020A&A...644A..55A}.
 For the present work we considered the diffusion of $\mathrm{H}$, $^{2}\mathrm{H}$ , $^{3}\mathrm{He}$
, $^{4}\mathrm{He}$ , $^{12}\mathrm{C}$, $^{13}\mathrm{C}$, $^{14}\mathrm{N}$, $^{16}\mathrm{O}$, and $^{22}\mathrm{Ne}$.
\subsection{Axion emission in white dwarfs}
\begin{table}[t]
\centering
\begin{tabular}{lcccc}
\hline
       &  $a_0$ &$a_1$&$a_2$&$a_3$\\
       & $b_0$ &$b_1$&$b_2$&$b_3$\\
\hline
\multicolumn{5}{l}{Helium } \\
 ($Z=2$)  & +0.413 & +0.233     & $-0.655$ & +0.058 \\
        & +0.931 & +0.096    & $-2.641$ & +1.226 \\
\hline
\multicolumn{5}{l}{Carbon } \\
 ($Z=6$)  & +0.248 & +0.306   & $-1.145$ & +0.393 \\
        & +1.293 & +0.152   & $-2.918$ & +1.200 \\
\hline
\multicolumn{5}{l}{Oxygen } \\
 ($Z=8$)  & +0.133 & +0.325   & $-1.155$ & +0.410 \\
        & +1.401 & +0.169   & $-3.006$ & +1.213 \\
\hline
\multicolumn{5}{l}{Neon } \\
($Z=10$)   & +0.016 & +0.335   & $-1.081$ & +0.384 \\
        & +1.495 & +0.184    & $-3.121$ & +1.248 \\
\hline
\end{tabular}
\caption{Coefficients for the fit functions of eqs.~(\ref{coef}).}
\label{coeftable}
\end{table}
 If axions are coupled to electrons, the dominant emission process in white dwarfs is electron-ion Bremsstrahlung~\cite{Raffelt1986, Krauss:1984gm, Raffelt:1985nk}. This process take place in a highly degenerate environment and is also characterized by strong ion-correlations ($\Gamma\gg 1$)\footnote{The ionic coupling constant ($\Gamma$) for multicomponent plasmas is defined  as 
 $$ \Gamma = 2.275 \times 10^{5} \frac{(\rho Y_e)^{1/3}}{T} \sum_i (n_i/n_{ions}) Z_i^{5/3}\ \ \ \hbox{[c.g.s.]},$$  
 where the sum is taken over all considered nuclear species.}. The treatment of Bremsstrahlung emission for such a regime has been accounted as in \cite{bottaro}, who have recently reviewed the bremsstrahlung emission of bosons in degenerate environments, including axions. Specifically, Bremsstrahlung axion emission was computed as 
 \begin{equation}
 \epsilon_{BD} = 10.85 \; \alpha_{26} \;  T_8^4\;  \sum_j^{N_{\text{isot}}} \frac{X_j Z_j^2}{A_j} \times F_j,
 \end{equation} \label{eq:emission}
 where $\alpha_{26}$ is the axionic fine structure constant\footnote{ $g_{ae}$ is also related to the axionic fine structure
constant $\alpha_{26}$ by $\alpha_{26} = 10^{26} × g_{ae}^2 /4\pi$.}, $T_8=T/10^8$ and $X_j$, $Z_j$ and $A_j$ are the mass fraction, atomic number and atomic weight of element $j$. The function $F_j$ is being computed as 
\begin{equation}
 F_{j}(x,\Gamma)=\mathcal{A}_j(x)\Gamma^{-0.37}+\mathcal{B}_j(x)\Gamma^{+0.03}    
\end{equation} 
in the case $1<\Gamma < 160$ and $4<\log_{10}\rho<7$, where $\rho$ is the density in units of g/$\text{cm}^{3}$, $x = \log_{10}(\rho)$, and 
\begin{equation}
\begin{aligned}
& \mathcal{A}_j(x) = a_o^j+a_1^j x+\frac{a_2^j}{(8-x)} + \frac{a_3^j}{(8-x)^2} \,
\\
& \mathcal{B}_j(x) = b_o^j+b_1^j x+\frac{b_2^j}{(8-x)} + \frac{b_3^j}{(8-x)^2} 
\end{aligned}
\label{coef}
\end{equation}
 are fit functions. These fitting functions are accurate at the few-percent level over their range
of applicability and improve on recipes used in earlier studies.
 Table  \ref{coeftable} shows the values of the coefficients computed by \cite{bottaro} for carbon and oxygen, together with the values for neon and helium that were computed specifically for the present work. On strong ion-correlation environments where the conditions established on \cite{bottaro} were not met we used the prescriptions derived in \cite{nakagawa1,nakagawa2}. 

The full set of axion recipes used in our code includes axion emission by Compton process, bremsstrahlung emission in the non-degenerate case as in \cite{Raffelt_book}, and at low ion-correlations in the degenerate case we used \cite{raffeltweiss}. For intermediate ion-correlations ($0.9 <\Gamma < 1.1$) 
 we interpolated linearly between recipes, and for intermediate degeneracy, which is of more relevance in stars from the Red Giant Branch (RGB) and the Asymptotic Giant Branch (AGB) than in white dwarfs \cite{carenzaylucente}, we interpolate with a harmonic expression as in \cite{miller2014}. 
\subsection{Initial models and white dwarf model grid}
\begin{figure}[t]
    \centering
    \includegraphics[width=6in,keepaspectratio]{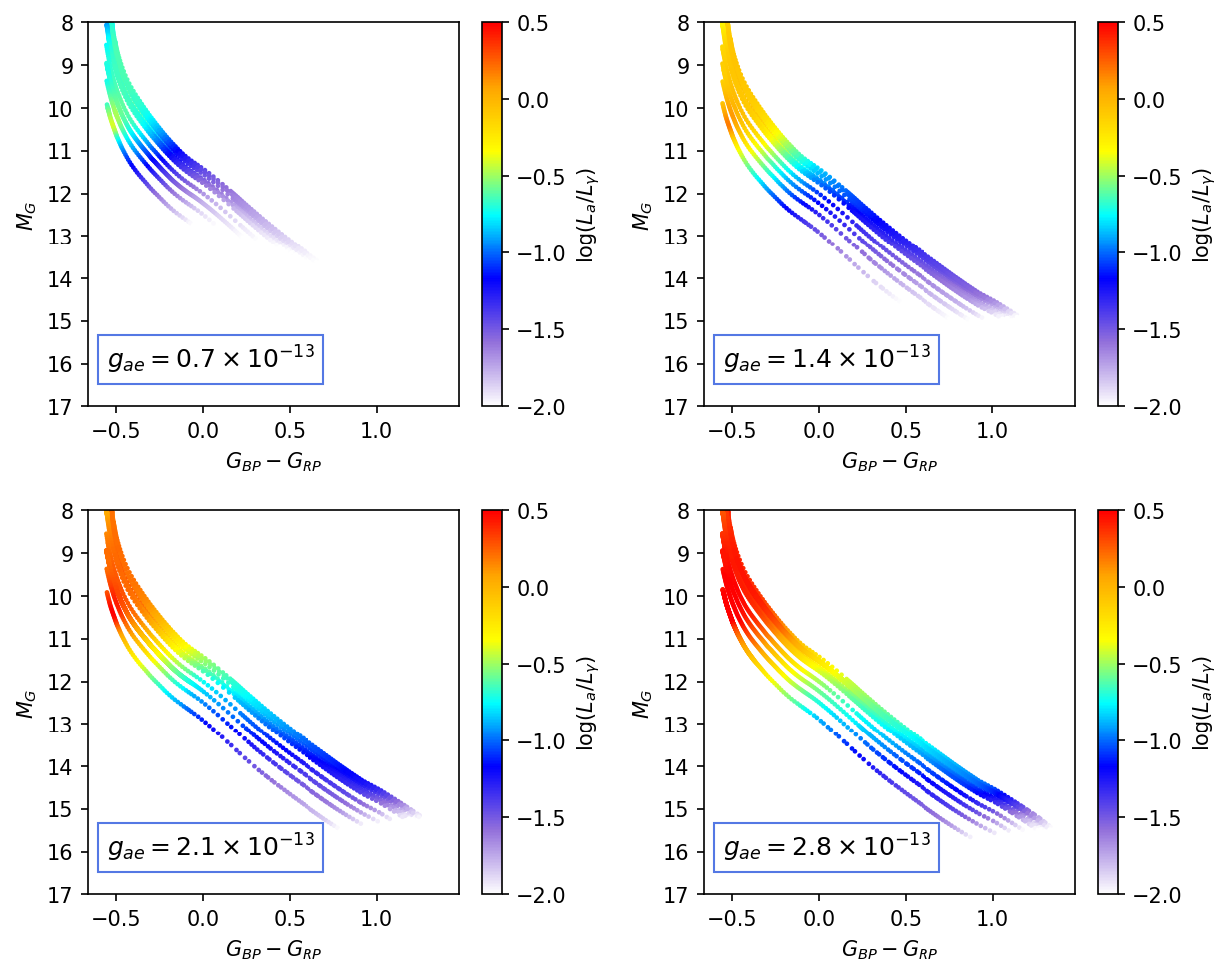}
    \caption{Evolutionary tracks in the color-magnitude diagram of white dwarfs with metallicity $Z = 0.01$, and masses in the range $0.48–1.05\; M_\odot$. Each panel considers different values of $g_{ae}$. The color scale shows the intensity of axion emission relative to the surface luminosity.}
    \label{ax-emisiv}
\end{figure}
The initial white dwarf models adopted in our simulations were taken from \cite{miller2016}. These models represent the initial states of the white dwarfs whose surfaces will be dominated by hydrogen, i.e. DA white dwarfs. The initial metallicities considered in these models were $Z = 0.0001$, $Z = 0.001$, $Z = 0.01$, and $Z = 0.02$. Interpolation between these values is sufficient to generate theoretical models for white dwarfs with metallicities typical of stellar clusters, the solar neighborhood, and even the Galactic halo.  The pre-white dwarf models were calculated in \cite{miller2016} for a set of given masses for the main-sequence progenitors, and are the result of the computation of the full previous evolution under the assumption of grounded wind prescriptions. Consequently, evolutionary computations do not yield neatly rounded white dwarf masses; rather, their values contain many decimals and differ for each metallicity. To improve the efficiency of interpolations between different evolutionary tracks in the Monte Carlo simulations, we rescaled the white dwarf masses---i.e., we changed the total mass of the white dwarfs keeping the chemical gradient fix in term of $m(r)/M_{\rm tot}$---constructing a grid of initial models with white dwarf masses of $0.48,0.50,0.53,0.56,0.58,0.60,0.70,0.80,0.90$ and $1.05 \,M_\odot$ for each metallicity. 
This procedure is justified because the net change in stellar mass before and after the rescaling is small, and the differences to the chemical stratification are negligible when compared with the current uncertainties in prior stellar evolution (e.g., the initial–to–final mass relation, which is sensitive to poorly constrained processes such as mass loss and convective overshooting). 
Then we calculated the white dwarf evolutionary models, considering several values of the coupling constant $g_{ae}$ between $0$ and $8.4 \times10^{-13}$, that translates, in terms of $m_a \cos^2 \beta$ under Eq.~\ref{eq:dfsz}, to: $0,\;2.5,\;5,\;7.5,\;10,\;12.5,\;15,\;20$ and $30$ meV. Therefore calculating 360 white dwarf models. Fig.~\ref{ax-emisiv} shows the evolutionary tracks for $Z = 0.01$ corresponding to the grid masses and for different values of the axion-electron coupling constant $g_{ae}$.
\begin{figure}
    \centering
    \includegraphics[width=1.\linewidth]{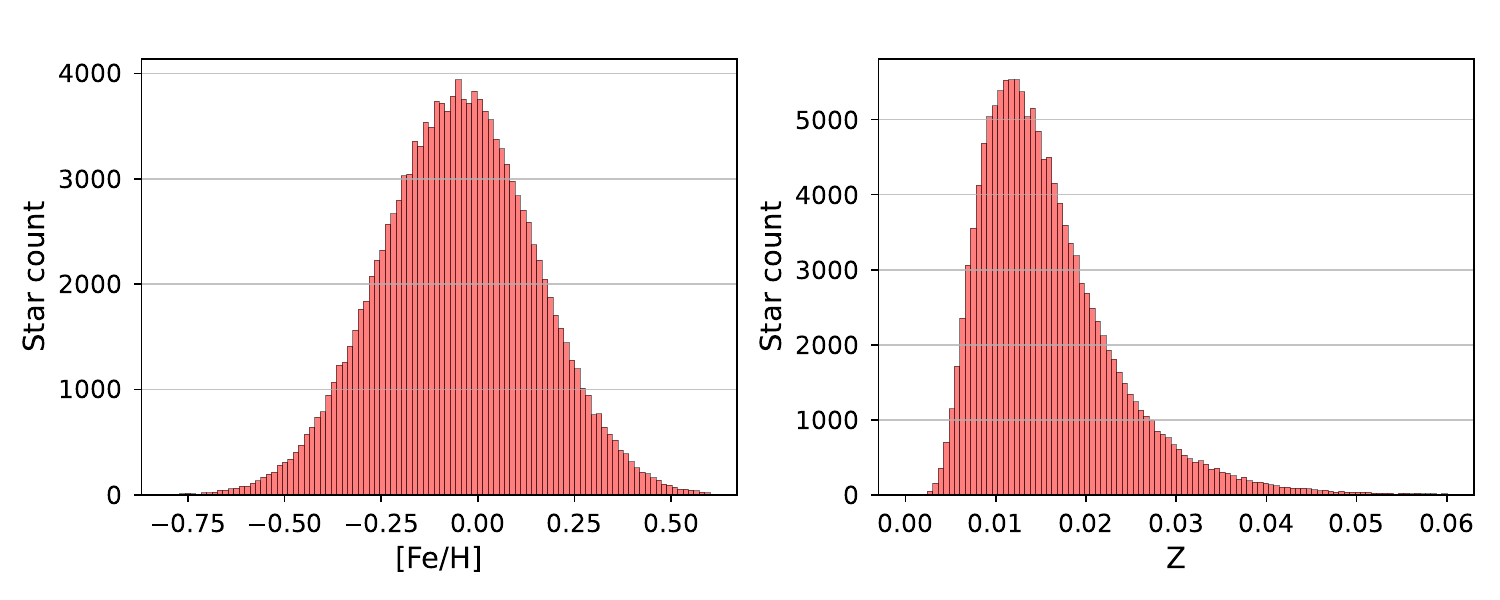}
    \caption{Left panel: Metallicity Distribution Function (MDF) of the solar neighborhood in terms of [Fe/H], in agreement with \cite{2011A&A...530A.138C}. Right panel: $Z$ distribution, related to [Fe/H]  via $Z = Z_0\;\cdot10^{\;\mathrm{[Fe/H]}}$, where we used $Z_0 = 0.016$ as the total solar metallicity, according to \cite{2025arXiv250305402B}.}
    \label{fig:met}
\end{figure}
\section{Construction of the white dwarf luminosity function} \label{sec:wdlf}
\subsection{Monte Carlo population synthesis code}\label{mcqub}
To construct the theoretical WDLFs for the Galactic disk we made use of the Monte Carlo population synthesis code \cite{Gberro1999} for synthetic white dwarf populations. This code has been widely used in the analysis of single \citep[e.g.][]{Torres2005,Torres2016,Jimenez2018,Torres2021} and binary \citep[e.g.][]{Camacho2014,Cojocaru2017,2018MNRAS.480.4519C,2022MNRAS.511.5462T}  populations of white dwarfs, as in studies of open and globular clusters \citep[e.g][]{2010Natur.465..194G,Torres2015} and the Galactic bulge \citep{Torres2018}. A detailed description of the code can be found in these references. Therefore, we will only review its basic inputs. 

Synthetic main sequence stars are generated with masses randomly following an initial mass function with a Salpeter distribution \footnote{The Salpeter initial mass function gives the number of stars $N(m)$ formed with masses in the range $(m,m+dm)$ as a power law of the form $N(m) dm \propto m^{-2.35} dm$ \cite{Salpeter1955}} and a minimum mass of $0.4 M_\odot$. Moreover, main sequence stars are generated at different initial times according to a given Stellar Formation Rate (SFR). Once every main-sequence star is generated, the BaSTI pre-white dwarf database \citep{2004ApJ...612..168P} is employed to determine which stars had the time to become WD. Then, an initial-to-final-mass relation (IFMR) is adopted to compute the resulting white dwarf masses and cooling times. In the present work, we used the IFMR presented in \cite{cummings}. 
To account for the Galactic disk metallicity, 
we used a bell-shaped curve for the Metallicity Distribution Function (MDF) of the solar neighborhood in terms of [Fe/H], in agreement with the one presented in   Fig.~15 of \cite{2011A&A...530A.138C} and Fig.~5 of  \cite{2021MNRAS.505.3165R}, related to $Z$ via $Z = Z_0\;\cdot10^{\;\mathrm{[Fe/H]}}$, where we used $Z_0 = 0.016$ as the total solar metallicity, according to \cite{2025arXiv250305402B}.
The resulting MDF and the corresponding distribution in $Z$ are shown in Fig.~\ref{fig:met}.
\begin{figure}[t]
    \centering
    \includegraphics[width=1.0\linewidth]{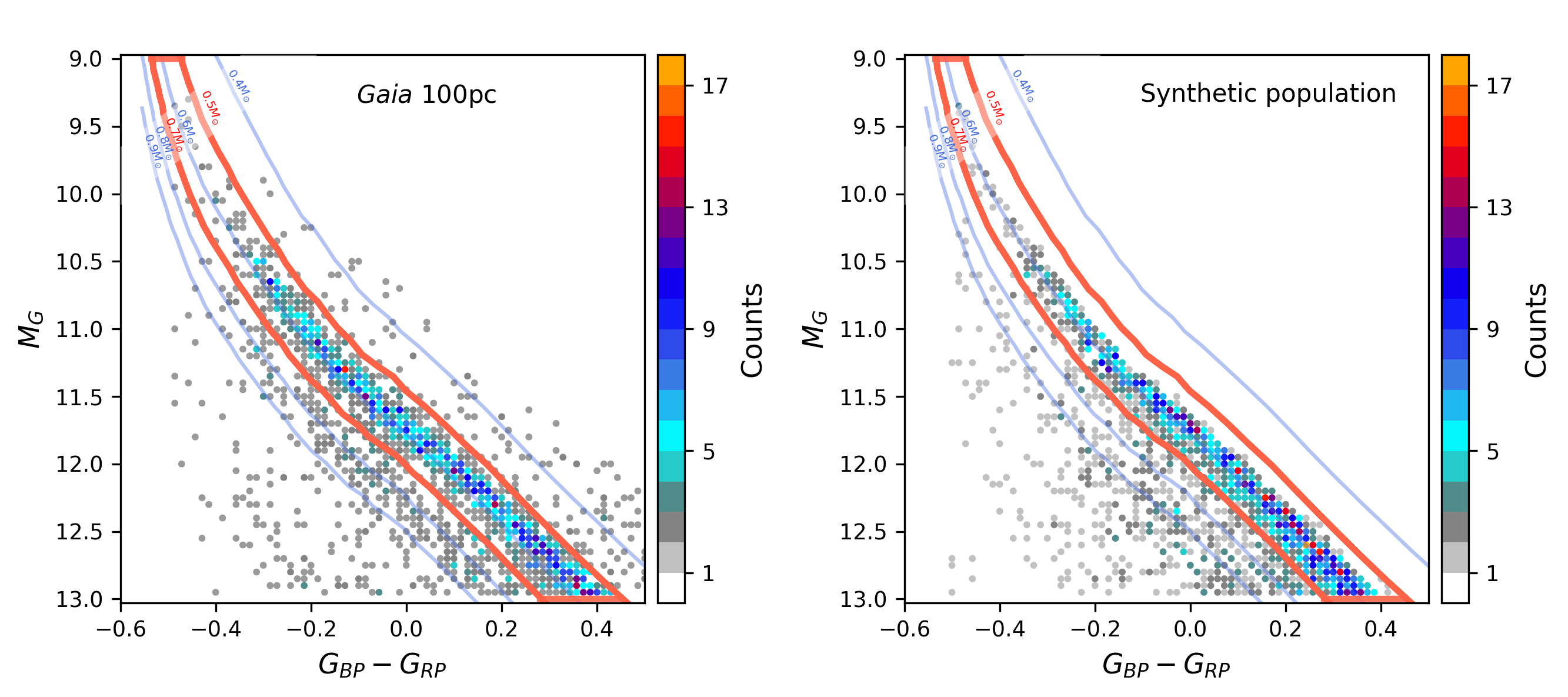}
    \caption{Hess diagrams for the {\em Gaia} 100-pc DA white dwarf sample from \cite{JE2023} (left panel), against a synthetic thin-disk DA white dwarf population of equal number of objects. Thin blue lines indicate the evolutionary tracks of white dwarfs of different masses. From top to bottom, 0.4, 0.5, 0.6, 0.7, 0.8 and 0.9 $M_\odot$. The red lines define the selected region of the color-magnitude diagram.}
    \label{gaiavssim}
\end{figure}

We considered that all white dwarfs have a thick H envelope, belonging to the DA spectral type.
To determine the current status of a white dwarf of given mass and cooling times, we interpolate between the precomputed white dwarf model grids to obtain luminosity, effective temperature, surface gravity, and radius of the WD.

Finally, in order to compare with the observational sample, the Monte Carlo code applies atmospheric models \cite{2010MmSAI..81..921K}
to convert the quantities of our stellar models into {\em Gaia} {\it passbands } magnitudes, and incorporates photometric and astrometric errors in concordance with {\em Gaia} performance \footnote{\url{http://www.cosmos.esa.int/web/gaia/science-performance}}.
Figure~\ref{gaiavssim} shows a comparison between the observed color–magnitude diagram and a theoretical counterpart constructed with the same number of objects as the {\em Gaia} sample. Absolute magnitudes $M_G$ correspond the standard {\em Gaia} broad $G$ band, while the color index index $G_{\rm BP}-G_{\rm RP}$ is computed from  Gaia's blue ($G_{\rm BP}$) and  red ($G_{\rm RP}$) photometric passbands.
Along the white-dwarf cooling sequence, decreasing $M_G$ values correspond to higher luminosity, while the color $G_{\rm BP}-G_{\rm RP}$ provides a proxy for the effective temperature (smaller values, indicate hotter objects). The colored density map (Hess diagram) encodes the number of objects per bin in this plane, and the thin blue curves indicate representative theoretical cooling tracks for different white-dwarf masses.
\subsection{Observed and theoretical white dwarf luminosity functions}
\begin{figure}[t]
\centering
\includegraphics[width=6in,keepaspectratio]{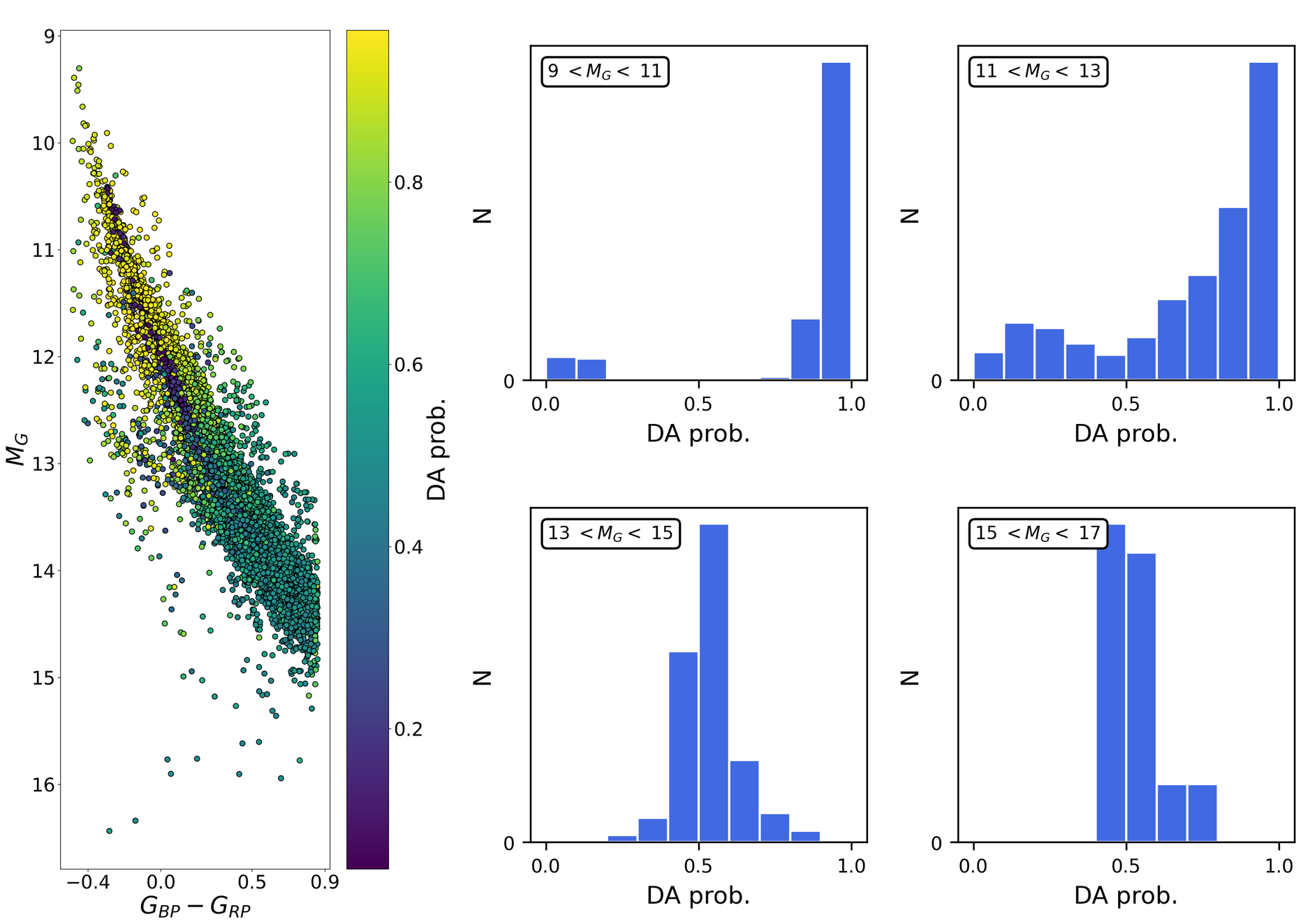}
\caption{left panel: the color-magnitude diagram of the 100 pc white dwarf population classified into DA and non-DA by   \cite{JE2023}. Color palette indicates the DA probability of each WD. Right panels: DA prob. distribution histograms for $9<M_G<11$, $11<M_G<13$, $13<M_G<15$, and $M_G>15$.  \label{histopda}}
\end{figure}
The observational sample that we will use to compare our theoretical simulations is the 100 pc {\em Gaia} DR3 white dwarfs sample from \cite{JE2023} 
, that belong to the Galactic disk according to \citep{2019MNRAS.485.5573T}. 
DA white dwarfs are better classified in the hotter regions of the color-magnitude diagram, where the spectral lines are more easily discernible (see \cite{JE2023}). Moreover, the axion emission is relevant on the brighter portion of the WDLF, as we can see in Fig.~\ref{ax-emisiv}. Using the  classification of~\cite{JE2023}, we analyzed the probability of a given star of belonging to the DA group. This probability is displayed in the color-magnitude diagram in the left panel of Fig.~\ref{histopda}. 
In the right panels of Fig.~\ref{histopda} we can see that the data exhibit a clear bimodality up until magnitudes $M_G < 13$, in which the two modes can be separated by a line at 0.5 in the DA probability, allowing us to statistically separate DA and non-DA white dwarfs.    
Beyond this magnitude, the spectral classification method is not good enough to tell DAs from non-DAs, as the probability distribution becomes unimodal. 
Alternatively, \cite{2023Garcia-Zamora} constructs a DA sample based on \cite{JE2023} using a Random Forest algorithm. We present our final results following each of these approaches in the construction of the observational sample.

The next step in this process is to generate a theoretical WDLF for the Galactic disk under the assumption of different intensities for the coupling constant $g_{ae}$. The synthetic population of white dwarfs is sensitive to the specific form of the SFR, which must be prescribed within the Monte Carlo code. In the next section, we briefly review our knowledge of the Galactic disk SFR, discuss its prescription, and outline the associated problems.

Given that we are restricting the modelling to CO-core DA white dwarfs coming from single stellar evolution, we need to minimize as much as possible the contribution from binary stellar evolution channels (post-merger and post-Roche lobe overflow white dwarfs), as well as those white dwarfs formed by Super AGB stars, in the observational sample. This can be achieved by excluding from the sample He-core white dwarfs, as well as massive CO-core  and ONe-core white dwarfs.
In light of this, when constructing the WDLF we have 
then restricted our observational sample to 
 those white dwarfs in the region delimited by the 0.5 and the 0.7 $M_\odot$ white dwarf evolutionary tracks, brighter than 13 $M_G$, and having  $P(DA)>0.5$.
The region of the color-magnitude diagram finally considered for the construction of the WDLF is delimited  by the red lines in Fig.~\ref{gaiavssim}. 
\subsection{The bright WDLF and its independence from the star formation history \label{sec:sfr}}
The SFR of the Galactic disk in the solar neighborhood is poorly constrained (see, for example, Fig.~9 of \cite{2025A&A...697A.128D}). This is not only due to our understanding of the Galactic stellar formation history, but also due to the complications that arise from stellar migration \cite{2024A&A...687A.286Z}. Previous works lead to discrepant results regarding the intensity and timing of stellar formation bursts within the Galactic disk, although they agree that the SFR has not varied by more than a factor 6 with bursts showing typical timescales of Gyrs.  Within this context, it is a reasonable assumption to adopt a constant SFR for the whole history of the Galactic disk, i.e. the SFR is adopted as a step function, where the step location is defined by the assumed age of the Galactic disk, and normalization is a free parameter. 
This SFR is not short of problems when compared to the observed WDLF, given that the later exhibits some evidence of bursts in the recent formation rate history \cite{mor2019,isern2019},  however, it involves a minimal set of assumptions and is therefore particularly suitable for the present work.
Here we have adopted a constant SFR for a Galactic disk of $10.6$ Gyr of age, consistent with \cite{cukanovaite}.
\begin{figure}[t]
\centering
\includegraphics[width=6in,keepaspectratio]{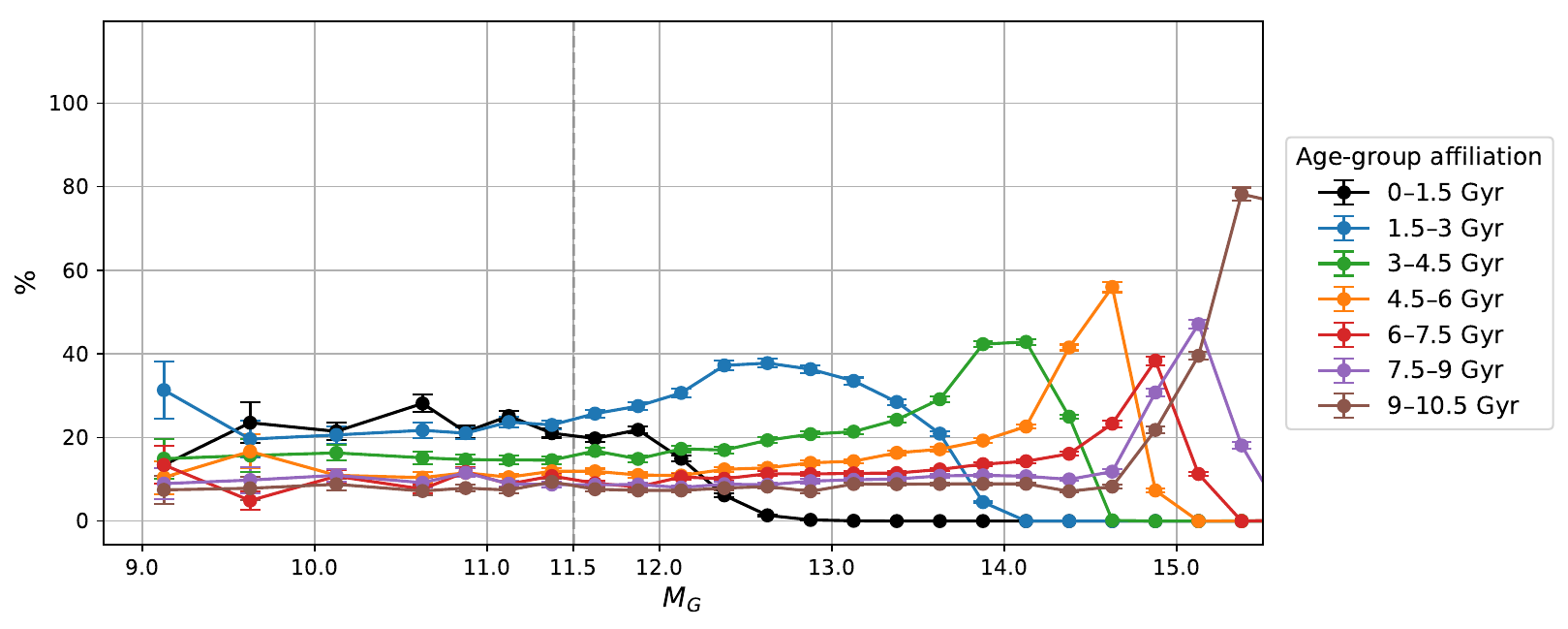}
\caption{Contribution of different stellar age bins (0–1.5 Gyr, 1.5–3 Gyr, 3–4.5 Gyr, 4.5–6 Gyr, 6–7.5 Gyr, 7.5-9 Gyr and 9-10.5 Gyr) to each magnitude bin in the WDLF.  \label{ageGroups}}
\end{figure}
Moreover, in the absence of very recent, short-lived bursts of star formation (shorter than 0.5 Gyr; see \cite{1990ApJ...352..605N}), for which there is no observational evidence \cite{2025A&A...697A.128D}, the brightest portion of the WDLF ($M_G\lesssim 11.5$) is expected to be largely independent of the SFR \cite{IsernGarciaBerro2008,Isern2009}, making this region particularly well suited for robust statistical tests on additional cooling channels, since the SFR constitutes our main source of uncertainty. In contrast, extending the analysis to fainter magnitudes, $M_G \gtrsim 11.5$, would be unsuitable for such statistical tests, given the current level of uncertainty in the SFR, as discussed in Appendix~\ref{sec:bursts}.

In order to test the independence of the theoretical WDLF---constructed here via Monte Carlo simulations--- from the SFR at the bright end, we 
 computed the fractional contribution of progenitors formed in seven stellar age bins ---each spanning 1.5 Gyr--- to each magnitude bin of the WDLF. Figure~\ref{ageGroups} shows the resulting distributions.
We find that up to $M_G \simeq 11.5$ the relative contributions of the different age bins remain nearly invariant within Poisson uncertainties, indicating that the bright WDLF is mostly insensitive to the SFR. 
Only at fainter magnitudes do progressively older stellar populations begin to dominate, and the age distributions cease to be invariant. And to further support this claim, in Appendix~\ref{sec:bursts} we evaluated the impact of six different young star formation bursts in the SFH. 
This numerical experiment therefore confirms that, in the bright region considered here and in the restricted white dwarf mass interval $0.5$--$0.7\,M_\odot$, the shape of the WDLF is governed by cooling physics, while the SFR affects only the overall normalization. This result is in line with the analytical demonstration found in \cite{Isern2009}.
\section{Impact of the axion emission in the WDLF} \label{sec:impact}
To assess the impact of axion emission on the shape of the WDLF, we computed theoretical WDLFs assuming different values of the axion–electron coupling constant $g_{ae}$. In line with the discussion in the previous sections, comparisons between the {\em Gaia} 100 pc WDLF and its theoretical counterparts are more robust when restricted to the region delimited by the 0.5 and 0.7 $M_\odot$ cooling tracks and to magnitudes $M_G < 11.5$.

In this luminosity regime, the WDLF is expected to be largely insensitive to fluctuations in the local SFR on Gyr timescales, as suggested by different observational determinations of the Galactic disk SFR. We have empirically tested this expectation by considering a wide range of SFR models, from constant histories to bursty scenarios with different ages. We then evaluated how different stellar age bins populate the WDLF as a function of $M_G$ (see Sec.~\ref{sec:sfr}). We found that the percentile relations between age bins remain nearly invariant up to $M_G \sim 11.5$, indicating that variations in the SFR have a negligible impact on the WDLF in this magnitude range.

Therefore, in what follows, we adopt WDLFs constructed under the assumption of a constant SFR, normalized to reproduce the theoretical WDLF in the range $9 < M_G < 11.5$.

\subsection{Binned methods}

The  $\chi^2$-tests for every dataset analyzed in this section was computed via Python's \href{https://docs.scipy.org/doc/scipy/reference/generated/scipy.stats.chisquare.html}{\textit{scipy} library}, which applies Pearson's  $\chi^2$-test as: 
\begin{equation}
\chi^2 = \sum_{i=1}^{N} \frac{(O_i - E_i)^2}{E_i}\;\; ; \,\; \nu = N-1,  
\end{equation}
where $E_i$ are the expected counts, $O_i$ the observed counts, $N$ is the total number of bins, and $\nu=N-1$ are the degrees of freedom, given that normalization is a fitted parameter. In order to properly estimate the theoretical expected values $E_i$, the theoretical realizations of the WDLF have been performed with a large sample of white dwarfs ($n\sim 10^5$). For each theoretical WDLF we have then computed the corresponding \textit{p-value}, i.e. the right-tail probability of observing a  $\chi^2$ value greater than the one calculated under the null hypothesis, 
\begin{equation}
 p\text{-value} = P(\chi^2\geq \chi^2_{\text{obs}}) .
\end{equation}

The theoretical WDLFs for different values of the axion-electron coupling constant $g_{ae}$ in the $9<M_G<11.5$ range display very good agreement with the observations when considering $g_{ae}=0$---i.e. no axion emission---and for values of $g_{ae}$ up to roughly $\sim 1.0\times 10^{-13}$. Beyond this point, the fits deteriorate rather abruptly. A comparison of the WDLFs for these models is shown in Fig.~\ref{fig:wdlfb} and the detailed statistical results are presented in Fig.~\ref{ajuste-full}. A table for the luminosity functions can be consulted at Tab.~\ref{tab:LFs}.
For the $\chi^2$-tests, 
several binning schemes were explored, all of them based on horizontal cuts in the magnitude axis, particularly binning schemes uniformly separated in $M_G$ and schemes were the bins had equal number of stars. 
All possible results from this schemes are well constrained by the results shown in Fig.~\ref{ajuste-full}, regardless of the fact that the figure exhibits only schemes with roundly equal number of stars per bin.

\subsection{Unbinned methods}
Due to the mild dependency of the statistical test on the adopted binning (Fig.~\ref{ajuste-full}) we performed additional  Kolmogorov-Smirnov (KS) tests (\href{https://docs.scipy.org/doc/scipy/reference/generated/scipy.stats.kstest.html}{\textit{scipy} library}) on the synthetic WD populations obtained under the assumptions of different values of $g_{ae}$.  

The KS test compares the cumulative distributions functions (CDF) along the $M_G$ magnitude of theoretical and synthetic WD samples following the same WD masses and magnitudes restrictions previously adopted in this section. The CDF for the \cite{zamora} WD sample against the CDFs for synthtetic populations calculated under different values $g_{ae}$ are shown in Fig.~\ref{fig:cdfs}. 
Notably, as shown in Fig.~\ref{ajuste-full}, the behavior of the p-value from the KS test is completely consistent with  those of our binned tests.

\subsection{Results}
As discussed above, both the binned and unbinned KS tests  are in very good agreement with each other. They all show a good agreement between models and observations  up to $g_{ae}\simeq 1.0\times 10^{-13}$  and a significant deterioration of the fit for larger values of the axion-electron coupling constant. Moreover values larger than $g_{ae}\simeq 3.0\times 10^{-13}$ are excluded by all tests at more than $99.7\%$ C.L. Due to its unbinned nature the KS tests behaves more smoothly and as a kind of consensus value among the different binned tests. For this reason we adopt this comparison as our main result. 
We conclude that values of $g_{ae}$ greater than $1.68 \times 10^{-13}$ can be discarded at $95\%$ C.L.~and values exceeding $2.52 \times 10^{-13}$ at the $99.7\%$ C.L. Note that this is true regardless of which of the two specific WD samples of the Galactic disk is adopted for the comparison. Results for both binned schemes and KS tests are shown in Fig.~\ref{ajuste-full}. In Table~\ref{ifmrstable} of Appendix~\ref{sec:ifmrs} we discuss alternative results obtained when considering two different IFMRs from the literature i.e. \cite{2026ApJ...996...69M} and \cite{2008A&A...477..213C}.

\begin{figure}[t]
\centering 
\includegraphics[width=1.\textwidth]{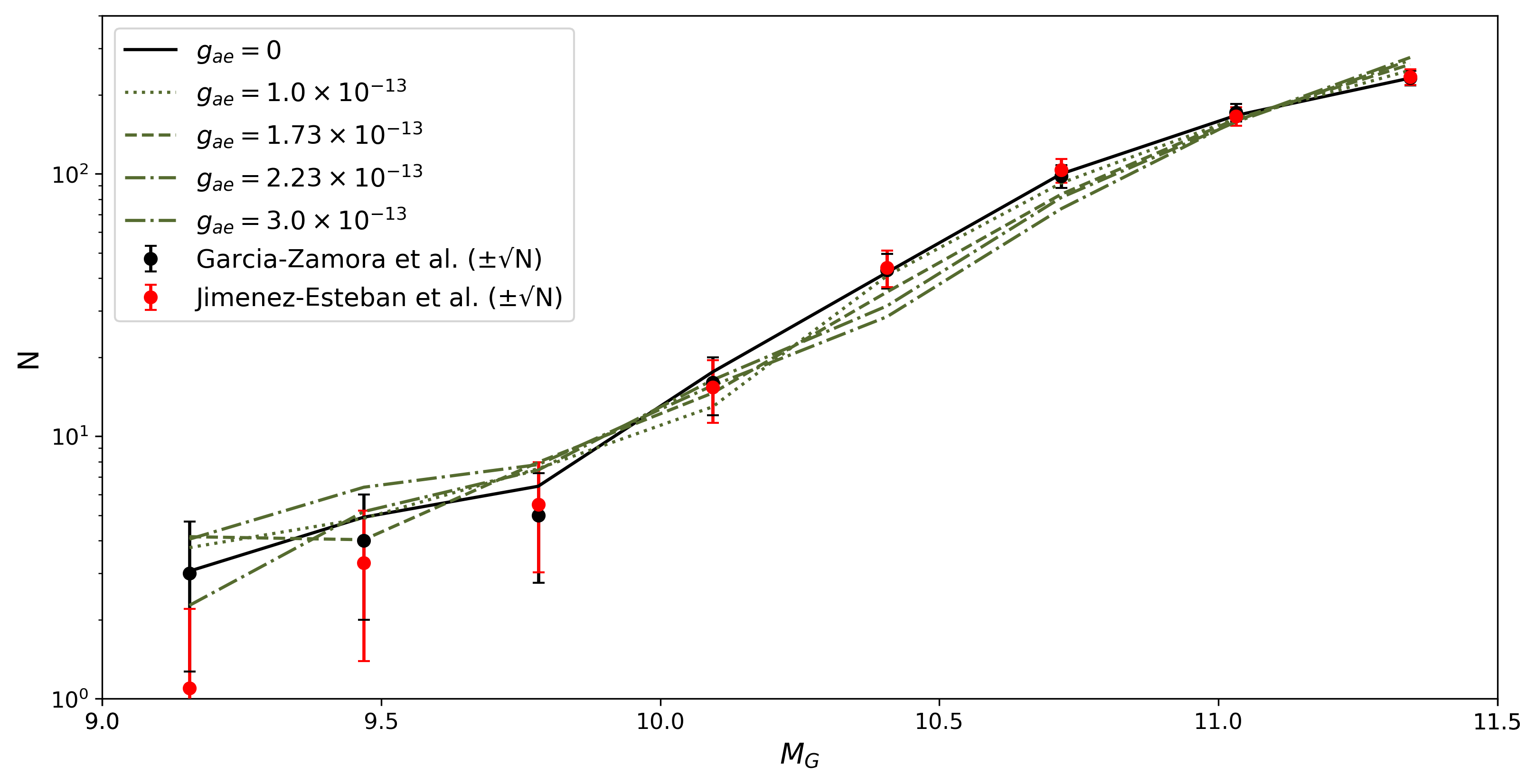}
\caption{WDLFs in the $9<M_G<11.5$ range for different values of $g_{ae}$. In black and red dots the observed values, using Garcia-Zamora et al.~DA sample \cite{2023Garcia-Zamora} and the Jimenez-Esteban et al.~\cite{JE2023} DA sample, respectively. In order to compare both samples in the same plot the counts in \cite{JE2023} where factorized by $N_{GZ}/N_{JE}$, with $N_{GZ}$ and $N_{JE}$ being the total number of WDs in the $9<M_G<11.5$ range for the \cite{2023Garcia-Zamora} and \cite{JE2023} samples, respectively. A table (Tab.~\ref{tab:LFs}) for the luminosity functions can be consulted at Appendix~\ref{app:LFS}. \label{fig:wdlfb}}
\end{figure} 

\subsection{Discussion}
\label{section:discussion}
The constraints on the electron-axion coupling constant derived in the present article $g_{ae}< 1.68 \times 10^{-13}$ 95\% C.L. and $g_{ae}<2.52 \times 10^{-13}$ 99.7\% C.L. supersede the previous constraints derived from the Galactic Disk by \cite{miller2014} from the pre-Gaia galactic disk luminosity functions.  The improvement is not only quantitative but in quality. Previous determinations by \cite{miller2014}  were based on incompatible data sets (see \cite{MillerBertolami2014}) as well as on simplified treatments of the WDLF that neglected observational and theoretical uncertainties. Therefore, besides the quantitative improvement on the constraint of the axion electron coupling constant the new constraints are  more robust and trustworthy.

The constraints on $g_{ae}$ presented in this work are very similar to those obtained by \cite{2020straniero}   from the RGB tip of 22 galactic globular clusters  $g_{ae}< 1.48 \times 10^{-13}$(95\% C.L.) as well as to the constraints derived from the RGB tip of the Galaxy NGC 4258 $g_{ae} < 1.48 \times 10^{-13}$ (95\% C.L., \cite{Capozzi:2020cbu}). Our constraints are, however,  weaker than the  $g_{ae} < 0.95 \times 10^{-13}$ (at 95\% CL) found when using only the 21 globular clusters with Gaia DR3 distances \cite{Carenza2025}. Although our current constraints are therefore not competitive with the latest RGB-tip bounds, they are relevant as they depend on different theoretical and observational systematics.

It should be noted, however, that our new constraints pale in comparison with the recent determinations based on galactic globular clusters,  $g_{ae} < 0.58 \times 10^{-13}$ (95\% C.L., \cite{2025Troitsky}), as well as with those derived by \cite{Fleury:2025ahw} based on the WD cooling sequence of 47 Tuc, $g_{ae} < 0.81 \times 10^{-13} $ (95\% C.L.). These two constraints however should be taken with skepticism. As  discussed in detail by \cite{Fleury:2025ahw}, the recent RGB-tip constraint presented by \cite{2025Troitsky} is strongly affected by the presence of two clusters that show a severe discrepancy with theoretical models and is significantly weakened to $g_{ae}< 1.2 \times 10^{-13}$(95\% C.L.) when those two clusters are excluded from the analysis. A similar issue lurks behind the constraints presented by  \cite{Fleury:2025ahw}. From the analysis presented in their appendix B (see their Tables 11 and 12) it is clear that their main constraint on $g_{ae}$ is driven predominantly by their WFC3 data, for which the authors' own goodness-of-fit KS diagnostics show statistically significant discrepancies between their data and their favored model  (of the order of $2-2.5\sigma$).  Thus, the strongest part of the constraint is being extracted from the dataset for which the adequacy of the adopted theoretical model is questionable. Moreover, the main constraint on $g_{ae}$ derived by \cite{Fleury:2025ahw} s also strongly influenced by their choice of priors. As shown in their appendix A, (e.g. see their Fig. 9 and Fig.20) when uniform priors are chosen on the total mass of the WD ($M_{\rm wd}$) and its envelope thickness ($q_H$) of the white dwarf (their “case 1” analysis, see their Table 5), the constraint on $g_{ae}$ is strongly weakened. The main constraint presented by \cite{Fleury:2025ahw}, corresponds to the so-called “case 3” which uses two key priors for $M_{\rm wd}$ and $q_H$ coming from \cite{Fleury2026MNRAS} This is not an inocuous choice. In fact the values of $M_{\rm wd}$ and $q_H$ in \cite{Fleury2026MNRAS} are derived, from the the same WD cooling models under the assumption of no axions (i.e. $g_{ae}=0$) and a very similar set of data. As the authors themselves acknowledge there is a degeneracy between the value of $g_{ae}$  and the envelope thickness, so including this prior information for the envelope thickness is important in their analysis for tightening the constraint on $g_{\rm ae}$. This is particularly concerning because the tightness of the constraint is therefore strongly dependent on the priors inherited from the authors' previous analysis of the same cluster using the same class of cooling models, data, and under the standard assumption of $g_{ae}=0$.

\section{Conclusions and future work} \label{sec:results}
We have derived new constraints for the DFSZ-axion using {\em Gaia} DR3 and white dwarf luminosity functions constructed with synthetic white dwarf populations through a Monte Carlo code. We integrated into the code 360 state-of-the-art white dwarf cooling sequences, where the axion emission was computed self-consistently with the thermal structure of the stars. 
We improve upon previous constraints on the axion–electron coupling constant derived from the WDLF in two main ways:

1. By constructing a theoretical WDLF with a Monte Carlo code (see Sec.~\ref{mcqub}), rather than relying on a semi-analytical approach, as seen in \cite{miller2014,isern2018}. This methodology allows us to account for {\em Gaia} observational biases  and selection criteria more effectively.

2. By exploiting the reliability of the {\em Gaia} DR3 100 pc volume-limited sample. In contrast, \cite{miller2014,isern2018} employed catalogs such as \cite{Rowellhambly2011,Harris2006}, which were constructed via proper motion methods and through magnitude-limited samples, therefore inheriting the uncertainties associated with these approaches in the determination of the WDLF.

Together, these improvements in both the theoretical modeling and the observational data set allow us to derive significantly more robust constraints than those obtained in earlier works.
\vspace{0.1cm}

\begin{figure}[htbp]
\centering
\includegraphics[width=1.\textwidth]{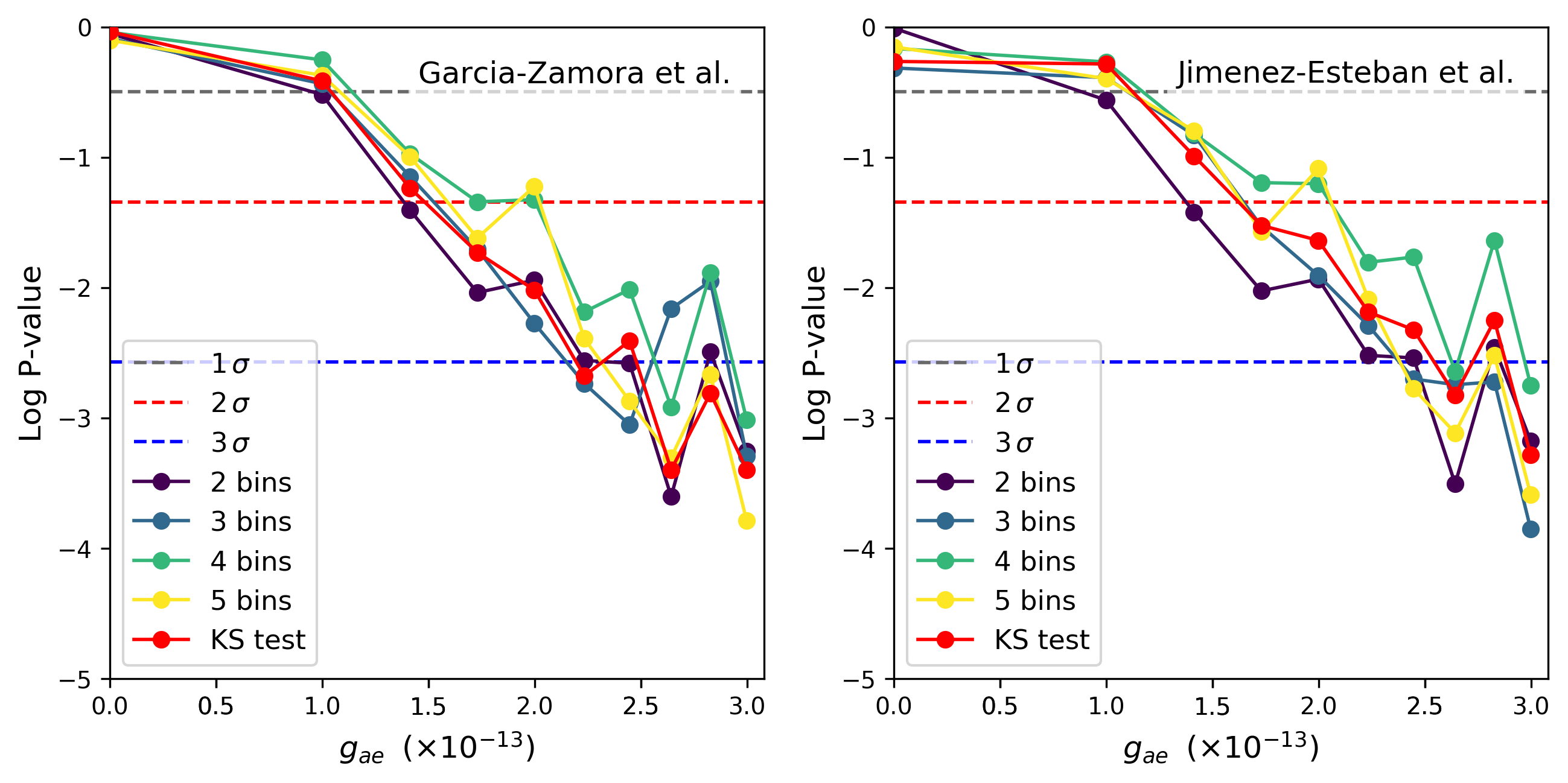}
 \caption{$p$-value as a function of $g_{ae}$ for the brightest white dwarfs ($9 < M_G < 11.5$), using the samples of \cite{JE2023} (left panel) and \cite{2023Garcia-Zamora} (right panel), and assuming a constant SFR. Different binning schemes are shown (2, 3, 4, and 5 bins), each constructed to contain roundly the same number of stars. The red and blue lines indicate the $2\sigma$ and $3\sigma$ confidence thresholds, respectively.
 \label{ajuste-full}}
\end{figure}

\begin{figure}[htbp]
\centering
\includegraphics[width=1.\textwidth]{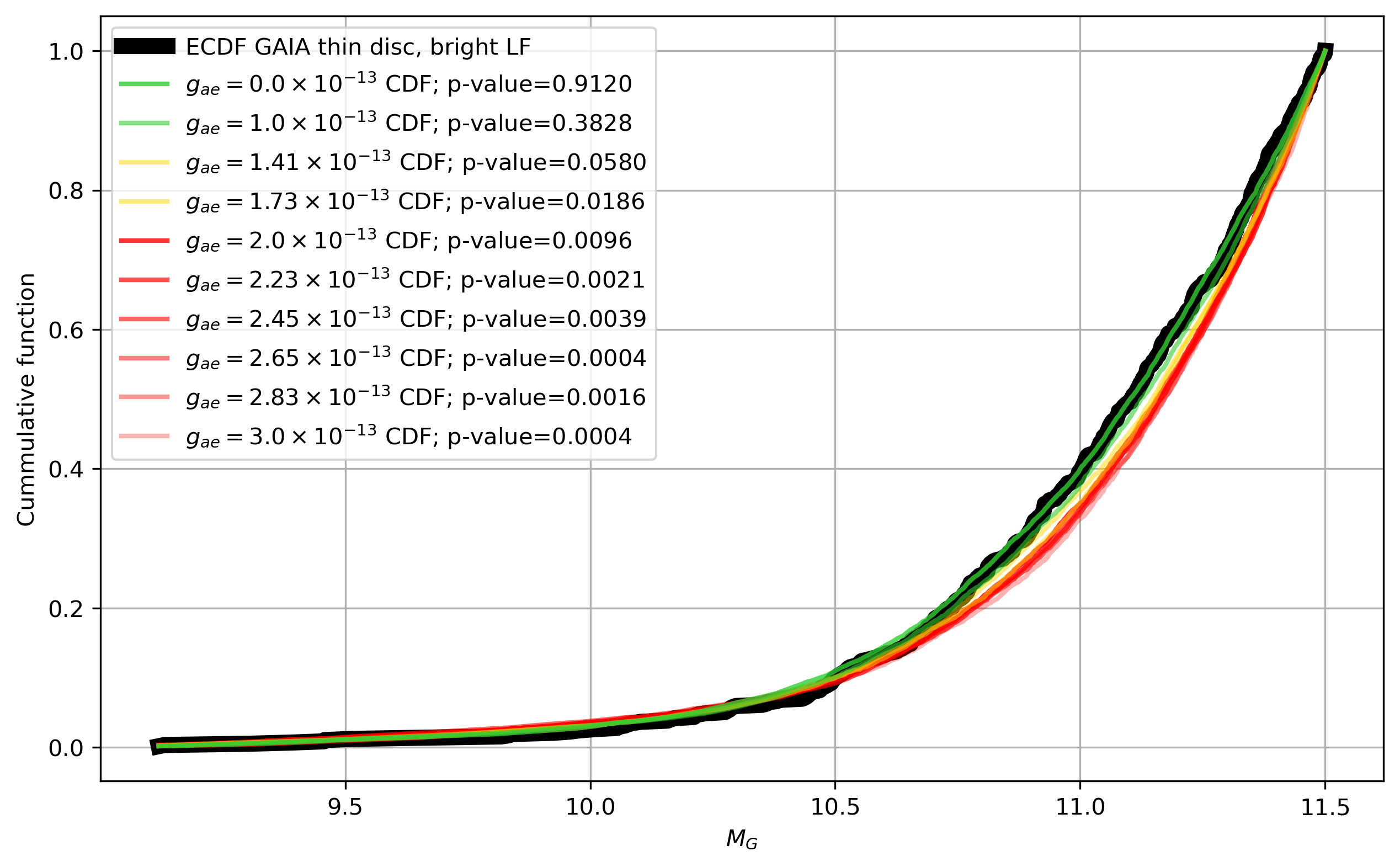}
 \caption{In black: CDF for the \cite{zamora} WD sample. In colors: CDFs under different values of $g_{ae}$, as depicted in the legend, along with their respective p-values under the KS test.
 \label{fig:cdfs}}
\end{figure}

Let us summarize our findings:

\vspace{0.2cm}
1.  The obtained $g_{ae}$ constraints were  $g_{ae} < 1.12 \times 10^{-13}$ at 68\% C.L.; $g_{ae} < 1.68 \times 10^{-13}$ at 95\% C.L.; and
 $g_{ae} < 2.52 \times 10^{-13}$ at 99.7\% C.L. (see Fig.~\ref{ajuste-full}). That translates, in terms of $m_a\cos^2\beta$ through Eq.~\ref{eq:dfsz}, to $4$, $6$, and $9$ meV, respectively. These constraints were obtained under a conservative but robust approach, restricting the analysis to the brighter portion of the sample, where the effects of the uncertainties in the SFR are almost negligible. 

2. Different binning schemes were explored, and in all the considered cases, the likeliness of the fits decreases relatively rapid as $g_{ae}$ becomes greater than  $\sim0.7 \times 10^{-13}$. As an unbinned counterpart, the KS test also provides an agreement with these results.

3. In order to have a better statistical test coming from Galactic disk observations a good understanding of the SFR is needed (see Appendix~\ref{sec:bursts}). A possible solution is to look for simpler populations. Recently, \cite{salaris2025} have presented a study of the white dwarf cooling sequence of the globular cluster 47 Tuc using deep infrared observations with the JWST, and estimating the cluster age as 11.8±0.5 Gyr. 

Besides the stellar formation rate, other limitations of this work include the neglect of mergers and binary evolution, and—on a more minor note—the adopted Galactic scale height, 
and the omission of axion effects on stellar thermal structure from the main sequence onward. In this context, we note that \cite{dominguez2025} report that ALP couplings can affect the estimated ages of massive CO white dwarf progenitors and modify the IFMR, potentially influencing population synthesis studies. However, these effects are not relevant for our purposes, as the disk ages explored here are substantially larger than the variations introduced by these ALP-induced corrections.

\begin{figure}
    \centering
    \includegraphics[width=1.0\linewidth]{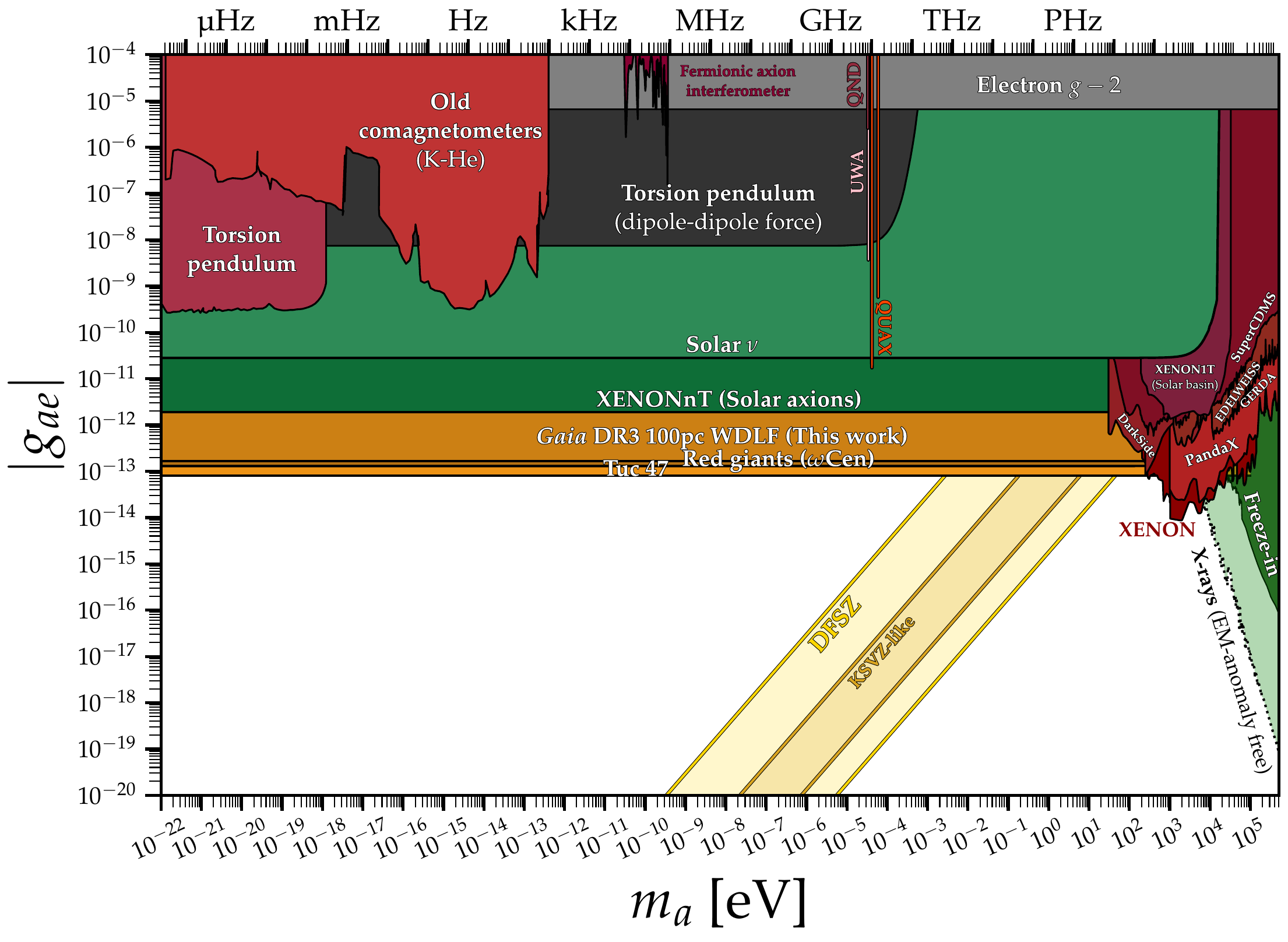}
    \caption{Constraints in the $g_{ae}$–$m_a$ plane from experimental and astrophysical observations (adapted from \cite{2020zndo...3932430O}). The orange-toned bands around $\sim 10^{-13}$ indicate the bounds derived from the {\em Gaia} DR3 WDLF (this work), red giants \cite{Capozzi:2020cbu}, and the recently obtained limits from Tuc~47 \cite{Fleury:2025ahw}. 
    }
    \label{fig:AxionElectron}
\end{figure}

As discussed in Section \ref{section:discussion} our bounds are comparable to those recently derived from the tip of the RGB  \cite{Capozzi:2020cbu,2020straniero}, and a factor 1.8 weaker than those derived from the RGB tip of 21 globular clusters with Gaia DR3 distances \cite{Carenza2025}. Our constraints are, however, significantly weaker than the most recent determinations by \cite{2025Troitsky} and \cite{Fleury:2025ahw}. As we discussed also Section \ref{section:discussion} we believe these constraints should be taken with a dose of healthy skepticism, due to dependence on several statistical assumptions and systematics. In any case,  the unavoidable presence of unknown unknowns is a key consideration when assessing the robustness of astrophysical constraints on particle physics. These issues highlight the complementarity between different approaches and the need for different, independent, astronomical observables. Fig.~\ref{fig:AxionElectron} presents these constraints in the $g_{ae}$–$m_a$ plane, along with other bounds obtained from experimental and astrophysical observations.

As for future work, in \cite{bottaro} Bremsstrahlung emission formulas are provided, not only for axions, but also for baryophilic and leptophilic scalars, under the assumption of degenerate, semi-relativistic electrons and ion–ion correlations in the liquid phase. In the case of scalars, however, the WDLF should depart from the standard one at large $M_G$, rather than small ones. We plan to extend our analysis to these additional particles and incorporate them into future studies of the WDLF in stellar clusters such as 47 Tuc \cite{salaris2025}.
\appendix
\section{Effects of stellar formation bursts on the WDLF} \label{sec:bursts}
\begin{figure}
    \centering
    \includegraphics[width=1.0\linewidth]{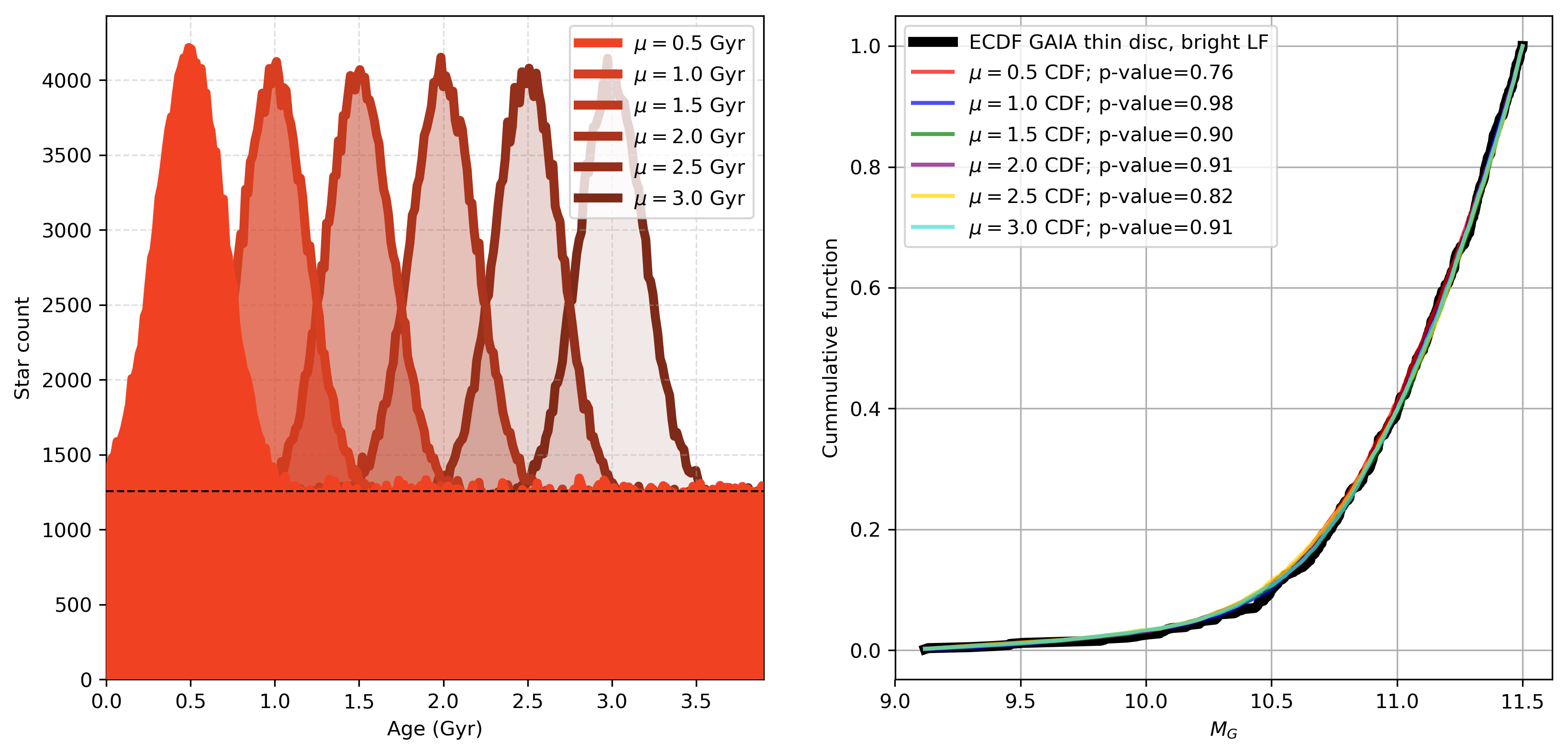}
    \caption{Left panel: Six SFR models consisting in constant components of 10.6 Gyr plus gaussian bursts of $\sigma=1$ Gyr and peaks at $\mu=$ 0.5, 1.0, 1.5, 2.0, 2.5, 3.0 Gyr, respectively.\\ Right panel: CDFs for the resulting synthetic populations under the previously described SFR bursts compared to the CDF for the Garcia-Zamora et al.~\cite{zamora} WD sample. WD magnitudes and masses are restricted according to the discussion in Sec.~\ref{sec:impact}. }
    \label{fig:youngbursts}
\end{figure}

In Sec.~\ref{sec:sfr}, we discussed why the brightest portion of the WDLF, $M_G \lesssim 11.5$, is particularly well suited for robust statistical tests of additional cooling channels, such as axion emission via electron–ion bremsstrahlung. 
To provide additional proof of this we performed KS tests under the assumption of different young bursts on the SFH (see Fig. \ref{fig:youngbursts}) consisting in a constant component of 10.6 Gyr plus gaussian bursts of $\sigma=1$ Gyr and peaks at $\mu=$ 0.5, 1.0, 1.5, 2.0, 2.5, 3.0 Gyr, respectively. We included the results of the KS tests on Table \ref{tab:ksbursts}. 

\begin{table}[htbp]
\centering
\begin{tabular}{lrccccccc}
\hline
& & KS test results \\
\hline
 Burst peak position (Gyr):  & 0.5 & 1.0 & 1.5 & 2.0 & 2.5 & 3.0 \\
\hline
p-value:    & 0.76 & 0.98 & 0.90 & 0.91 & 0.82 & 0.91    \\
\hline
\end{tabular}
\caption{p-values for the KS tests for the bright LF (see Sec. \ref{sec:impact} for def.) when considering different bursts in the SFH (see Fig. \ref{fig:youngbursts}). } \label{tab:ksbursts}
\end{table}

The same cannot be said when the analysis includes WDs dimmer than $M_G \sim 11.5$.

\begin{figure}[t]
\includegraphics[width=1.\textwidth]{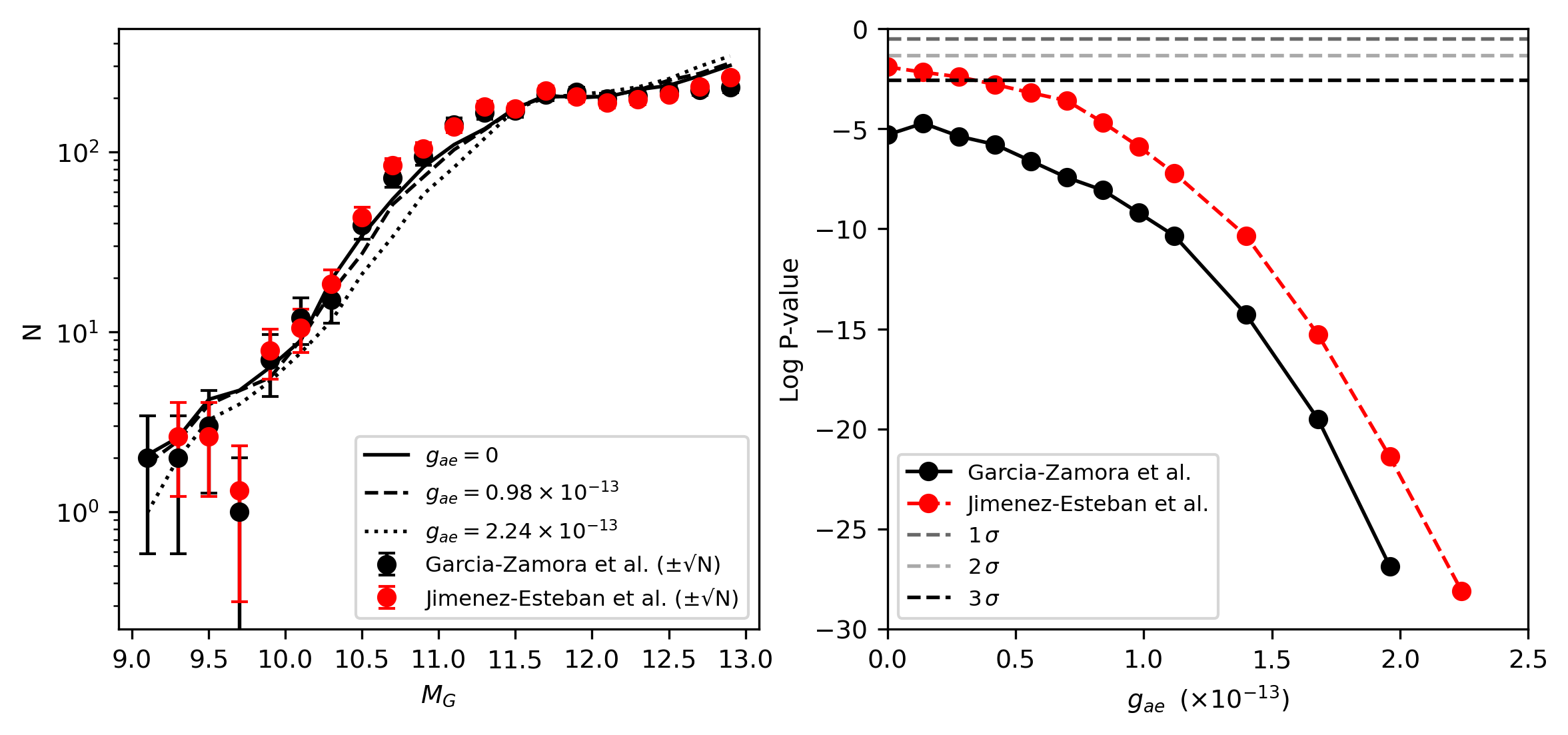}
\caption{Left panel: with lines, the theoretical WDLFs on the range  $9<M_G<13$, for different values of $g_{ae}$, using a constant SFR. Right panel: the $p$-values in function of the coupling constant $g_{ae}$.\label{fig:full-pvalor}}
\end{figure}

For instance, in Fig.~\ref{fig:full-pvalor} we show the WDLFs in the range $9 < M_G < 13$ for different values of $g_{ae}$, assuming a constant SFR, together with their corresponding $p$-values. Although the fits systematically worsen for larger values of $g_{ae}$, none of the tests yields a good fit. Normalizing the counts over this magnitude range leads to an underestimation of the WDLF at $M_G \lesssim 11.5$, followed by an overestimation at the fainter end.

Reproducing such trends would require white dwarfs to cool more slowly in the bright part of the diagram, opposite to the effect expected from additional anomalous cooling due to axions. A more plausible explanation is that the shape of the observed WDLF can be reproduced by a localized enhancement of the SFR affecting the progenitors of WDs populating the brighter region of the color–magnitude diagram (see Fig.~\ref{ageGroups}). As discussed in Sec.~\ref{sec:sfr}, sufficiently long-lived bursts do not affect the slope of the bright part of the WDLF ($M_G \lesssim 11.5$), but they increase the total number of WDs in that region and flatten the slope of the WDLF at fainter magnitudes ($M_G \sim 13$), thus improving the agreement with the data.

\begin{figure}[t]
 \includegraphics[width=1.\textwidth]{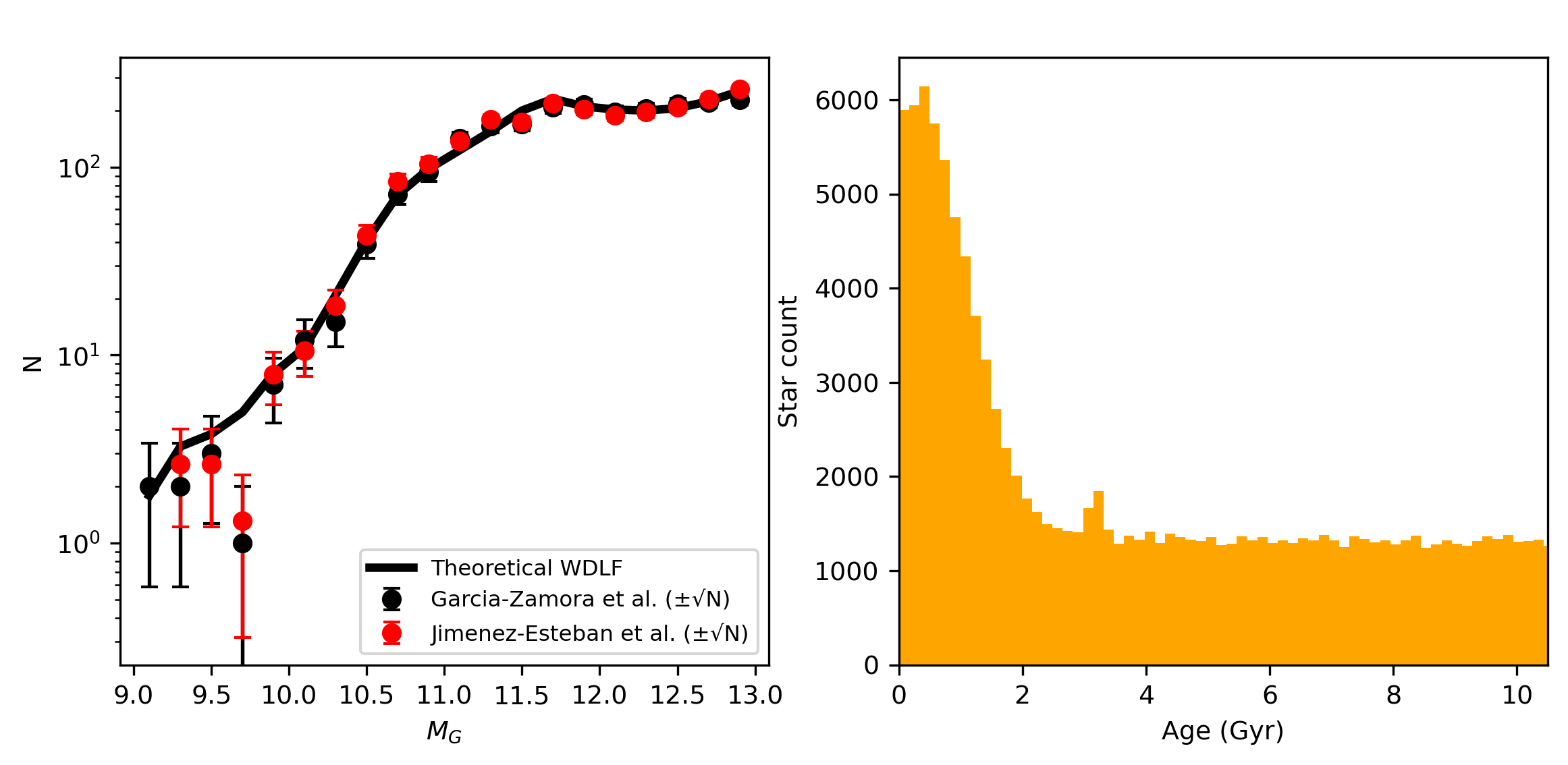}
 \caption{Left panel: WDLF adjustment using a constant SFR over 10.6 Gyr plus two Gaussian bursts, resulting in a $p$-value $\gtrsim74\%$. Right panel: the adjusted SFR.\label{fig:adjustment}}
\end{figure}

Indeed, we find that it is possible to reproduce the observed WDLF morphology by adopting a SFR with additional degrees of freedom. In particular, we obtain acceptable fits ($p$-value $\gtrsim 74$) for a model consisting of a constant SFR over 10.6 Gyr plus two Gaussian bursts (see Fig.~\ref{fig:adjustment}). However, this model introduces several free parameters, including the mean age, width, and normalization of each burst, as well as the normalization of the constant component. This highlights a fundamental limitation of this approach: In order to use the WDLF to robustly constrain the axion–electron coupling, the SFR must be determined independently of the white dwarf sample; otherwise, the additional free parameters could be used to artificially improve the quality of the fits, biasing the results toward certain values of $g_{ae}$. Given the current uncertainties in the observational determination of the SFR in the solar neighborhood, this requirement appears difficult to satisfy.

\section{Impact of different IFMRs on the bounds} \label{sec:ifmrs} 

We explored the effects of considering different IFMR from the literature, i.e. Miller et al. 2026 \cite{2026ApJ...996...69M} and Catalan et al. 2008 \cite{2008A&A...477..213C}. See Table \ref{ifmrstable}. The results shows that the 95\% C.L bounds are roughly the same as when considering Cummings et al. IFMR \cite{cummings}.

\begin{table}[htbp]
\centering
\begin{tabular}{lrcc}
\hline
& & KS test results (p-value) \\
\hline
  & IFMR Model: &  Miller et al. 2026 & Catalan et al. 2008\\
\hline
$g_{ae}$ ($\times 10^{-13}$) & & &  \\

$0$ &   & 0.83 & 0.73 \\
$1.68$ &  & 0.036 & 0.1 \\
\hline
$g_{ae}$ ($\times 10^{-13}$) \\
 bounds at 95\% C.L.:  & & 1.65 & 1.8 \\ 
\hline
& & 
\end{tabular}
\caption{p-values for the KS tests for the bright LF (see Sec. \ref{sec:impact} for def.) when considering different
alternatives to the IFMR coming from the literature.\label{ifmrstable}}
\end{table}

\section{Luminosity functions for the thin disc and synthetic populations at bright magnitudes} \label{app:LFS}

We present a table (See Tab. \ref{tab:LFs}) with the LF of Fig. \ref{fig:wdlfb}. 

\begin{table}[htbp]
\centering
\begin{tabular}{l|r|c|c|c|c|c}
\hline
 $M_G$  & $<M_{bol}>$ & $n_{GZ}$ & $n_{JE} \times \frac{N_{GZ}}{N_{JE}}$ & $n_{gae=0}$ & $n_{gae=1.72}$ & $n_{gae=3.0}$  \\
\hline
9.16 & 5.4 & 3 & 1.1 & 3.06 & 4.1 & 4.1 \\
9.47 & 6.1 & 4 & 3.3 & 4.9 & 4.0 & 6.4  \\
9.78 & 6.7 & 5 & 5.5 & 6.4 & 8.0 & 7.8 \\
10.09 & 7.3 & 16 & 15.4 & 17.6 & 14.6 & 15.6  \\
10.4  & 8.0 & 43 & 43.9 & 42.1 & 35.4 & 28.5  \\
10.72 & 8.7 & 98 & 103.2 & 99.9 & 83.6 & 73.6  \\
11.03 & 9.4 & 171 & 165.8 & 166.7 & 161.4 & 158.8  \\
11.34 & 10.2 & 232 & 233.9 & 231.4 & 260.9 & 277.3  \\
\hline
\end{tabular}
\caption{ Luminosity functions of DA white dwarfs shown in Fig. \ref{fig:wdlfb}. The Column $M_G$ represents each bin center, $<M_{bol}>$ 
is an estimate of the overall bolometric magnitude (note that this value would vary for white dwarfs of different masses and here we have restricted the samples to range $0.5<M/M_\odot<0.7$). $n_{GZ}$ and $n_{JE} \times \frac{N_{GZ}}{N_{JE}}$ are the bin counts for the Garcia-Zamora and renormalized Jimenez-Esteban samples, respectively. $n_{gae}$ represents the counts for the normalized synthetic populations when considering the said value of $g_{ae}$.}\label{tab:LFs}
\end{table}

\acknowledgments
MLA \,and\, MMMB\, were \,supported\, by\, CONICET\, and \,Agencia\, I\,+\,D+\,i \,through\, grants \\ 
PIP-2971 and PICT 2020-03316. MEC acknowledges 
grant RYC2021-032721-I, funded by 
MCIN/AEI/10.13039/501100011033.  MEC \,and \,ST \,were \,supported \,by \,the 
\, European \,UnionNextGeneration \, EU/PRTR, \, the \,
AGAUR/Generalitat \, de \, Catalunya \, 
grant SGR-386/2021, and by the
Spanish MINECO grant PID2023-148661NB-I00.
AC is supported by an ERC STG grant
(``AstroDarkLS'', grant No. 101117510).
AC acknowledges the Weizmann
Institute of Science for hospitality at
different  stages of this project and the support from the Benoziyo Endowment Fund
for the Advancement of Science. We acknowledge Jordi Isern and Georg Raffelt for fruitful discussions and comments on the original version of the manuscript that helped us to improve the  presentation and content of the article.
\bibliographystyle{JHEP}
\bibliography{biblio}
\end{document}